# Feasibility Assessment of Multitasking in MRI Neuroimaging Analysis: Tissue Segmentation, Cross-Modality Conversion and Bias correction

Mohammad Eslami[1,2], Solale Tabarestani[2], Malek Adjouadi[2]

[1]Massachusetts Eye and Ear, Harvard Medical School, Boston, Massachusetts, USA
[2] Department of Electrical and Computer Engineering, Florida International University, Miami, Florida, USA

Neuroimaging is essential in brain studies for the diagnosis and identification of disease, structure, and function of the brain in its healthy and disease states. In medical image analysis, neural networks and deep learning (DL) techniques are delivering promising results. Literature shows that there are advantages of multitasking with some DL schemes in challenging neuroimaging applications. This study examines the feasibility of using multitasking in three different applications, including tissue segmentation, cross-modality conversion, and bias-field correction. These applications reflect five different scenarios in which multitasking is explored and 280 training and testing sessions conducted for empirical evaluations. Two well-known networks, U-Net as a well-known convolutional neural network architecture, and a closed architecture based on the conditional generative adversarial network are implemented. The data used is from the MRBrainS18 challenge dataset and the leave-one-subject-out cross-validation scheme is addressed. The challenge dataset used for this research endeavor includes the multimodal images, as well as segmentation masks. For bias correction and contamination, 8 different bias fields are added to simulate the contaminated images. Different metrics such as the normalized cross-correlation coefficient and Dice scores are used for comparison of methods and results of the different experiments. Statistical analysis is also provided by paired t-test. The present study explores the pros and cons of these methods and their practical impacts on multitasking in different implementation scenarios. This investigation shows that bias correction and cross-modality conversion applications are significantly easier than the segmentation application, and having multitasking with segmentation is not reasonable if one of them is identified as the main target application. However, when the main application is the segmentation of tissues, multitasking with cross-modality conversion is beneficial, especially for the U-net architecture.

**Keywords** Neuroimaging Analysis, Segmentation, Bias-field Correction, Cross-modality Conversion, Deep learning

## A. *INTRODUCTION*

Magnetic resonance imaging (MRI) is an important recording modality for a wide variety of diseases, and is an effective imaging platform for both clinical and research trials. Neuroimaging analysis is strongly influenced by MRI scans in view of its diffusion properties that enables visualizing brain structure with the ability for detecting brain abnormalities and extracting different structural features including size, shape, and the localization of the different brain regions [1]. Accurrate localization of diseased regions could lead to targeted interventions and treatment protocols. Moreover, MRI has become a standard imaging modality due to its reduced exposure to radiation in comparison with x-rays for example.

MRI variations are important in rendering an accurate diagnosis and help guide treatment decisions since different MRI modalities have different characteristics for representing brain structure and related abnormalities. The T1 and T2-weighted scans are the most commonly used MRI sequences for structural image-based diagnosis and analysis. However, MRI procedure, despite being non-invasive, is a lengthy and expensive process, which makes it almost impossible to obtain all different subtle variations regardless of the level of precision [2]. This highlights the importance of developing sophisticated algorithms capable of converting between these various modalities without any added cost. In addition, modality-conversion is essential for many medical image registration applications [3], especially when the aim is to associate brain function to the structure or anatomy of a given brain region.

Field intensity inhomogeneities are intrinsic issues to MR imaging modality due to wide variations in the RF field, inhomogeneities in the static field, and gradient disturbance of eddy current, to name a few. These illumination artifacts or bias field, which is a smooth low-frequency disturbance which significantly degrades the edges or contours of the high-frequency details [4]. The degradation of image quality impacts the diagnosis and affects the results of all computer-based algorithms that rely on spatial invariance or multimodal imaging with one modality being degraded. In view of this fact, great efforts has been made toward developing effective methods and algorithms to attenuate and correct the bias field [5][6].

Image segmentation is well-known in many computer-aided methods, including neuroimaging analysis, in which the brain's anatomical structures are visualized and measured through extremely challenging segmentation tasks albeit relying still on atlases. Additionally, segmentation is useful for analyzing subtle brain changes in disease prone areas or in specific 3D sources for surgical planning and image-guided interventions or therapeutic protocols. Many algorithms have been developed to address this problem, including those that exploit deep learning (DL) and machine learning (ML) tools [7][8].

Recent retrospectives and surveys on the merits of DL and ML to analyze medical images have shown promising results and findings as well as the pitfalls that need to be overcome [9][10]. The need to have access to quality variations of MRI modalities to be processed via segmentation algorithms while confronting the bias field, drew the attention for establishing data processing pipelines that can partially or completely automate these image restoration tasks [11]-[15]. Multitask learning as proposed in this study is a machine learning approach that solves multiple tasks at the same time by exploiting similarities between tasks, consolidating their strengths and interrelatedness, and adding key constraints to the system to enhance the performance of a desired task. The aim is thus to search for a common set of features or attributes in between tasks to enable all tasks to be trained as one multitask model. This enables the performance of the model to be improved at an expedited rate. Xu et al., presents an algorithm for MRI image segmentation with anti-noise and bias correction, based on point-to-plane algebraic distance constraints [11]. Zhang et al., also proposed an approach for segmentation and bias correction of MRI brain images with inhomogeneity and noise through a novel and robust clustering method using contextual information [12]. The authors in [13] proposed an active contours approach to perform MRI image segmentation and bias field correction. This multi-phase active contour method is based on a clustering approach that can perform both tasks precisely. Another recent paper in this domain is [14], which was also devoted to accomplishing segmentation and bias field correction simultaneously by integrating expectation-maximization and a hidden Markov random field framework. In the area of multitasking, researchers in [15] explored the possibility of generating high-resolution imagery from low-resolution mono-modal images. Their approach was based on a weakly supervised convolutional sparse coding algorithm to simultaneously address the cross-modality image synthesis and super-resolution problem. Huang et al., proposed an unsupervised multi-task generative adversarial network for modality transfering and brain MRI image synthesis [16].

The focus of this research was on the potential of multitasking to include in one setting brain tissue segmentation, cross-modality conversion, and bias field correction. Two well-known deep learning methods are selected for this study. The *U-Net* network [17][18] and the *pix2pix* network [19]. Previously, we have investigated the performance of both networks in single-task and multitask fashions in previous works for chest X-ray image analysis (segmentation and bone suppression) [20] and low-dose CT image analysis (kidney segmentation and image enhancement) [21]. We shared the source codes and details of the hyperparameters on GitHub (https://github.com/mohaEs/image-to-images-translation). The rest of the paper is organized as follows. Section 2 described the methods and materials. Section three reports on the experimental settings and results. Section 4 provides a discussion and retrospective on the results.

## *B. Methods and Materials*

Two well-known deep learning methods are selected for this study. The *U-Net* network and the *pix2pix* network. *U-Net* or Unet is an autoencoder shaped network with skip connections. The *pix2pix* is a conditional generative adversarial network (*cGAN*) having a generator network for producing the desired outputs and a discriminator network for determining if the generator’s outputs are real or fake. The ideal generator is one that produces outputs close to the target images, fooling as a consequence the discriminator to assume that such outputs are the real ones. Both *U-Net* and *cGAN* methods are implemented for both single tasks and multitask fashions for a more comprehensive evaluation. For fair comparison, the network structure of the *U-Net* and the *cGAN*’s generator are selected to be the same. Both network types are capable of analyzing 2D frames without sacrificing generality.

The MRBrainS18 challenge dataset has been selected for this study due to its multimodal registered neuroimaging volumes with segmentation masks [22]. *MRBrainS18* contains multimodal brain volumes (T1, T2-FLAIR, and T1-Inverse) with segmentation labels of gray matter, white matter, cerebrospinal fluid, and other tissues on 3T MRI scans of the brain of seven subjects. Each volume contains 40 axial slices with size 240×240. Each slice is augmented four times with different rotations, zooming, and translation similar to [20]. For simulating the bias corruptions, the contaminated slices are produced by adding 8 different low-frequency bias fields as in reference [5].

T1 and T2-Flair modalities are selected as inputs and 5 different scenarios of experiments are considered to investigate the feasibility of multitasking in neuroimaging analysis (10 experiments in total):

1. Cross-modality conversion.
2. Bias correction & cross-modality conversion.
3. Tissue segmentation (gray matter, white matter, CSF) & cross-modality conversion.
4. Tissue segmentation on contaminated slices & bias correction.
5. Tissue segmentation on contaminated slices & cross-modality conversion.

Intensity of all volumes is stretched between 0 and 255 and each slice is resized to 512×512. The leave-one-subject-out (LOSO) cross-validation (CV) scheme is assumed. This research is thus based on 280 training and testing sessions (i.e., 2(input modalities) × 5(scenarios) × 7(LOSO-CV) × 4 (methods)).

### C. ***Results***

The network has been implemented using Python and the Tensorflow library. All computations for training the network have been performed using the NVIDIA DGX-1 platform (1 GPU v100 with 24GB RAM used). The learning rate of the Adam optimizer is set to 0.0002 and the kernel size of filters is 3×3. The batch size is 20 and the maximum epoch is 500 (force stop). The training process of both methods is stopped when the L1 loss value reaches 0.01 (early stop). For the cGAN network l1-weight is set to 10 (see [20]) and the training is also stopped if the loss of discriminator decreases while the loss of generator grows up for 10 consequent epochs (force stop). We have shared the source codes and details of all hyperparameters on GitHub (https://github.com/mohaEs/image-to-images-translation). In the experiments, ST and MT stand for single task and multitask implementations. The trainable parameters of methods are 74,750,703, 76,292,808, 74,756,106 and 76,299,939 for Unet-ST, cGAN-ST, Unet-MT and cGAN-MT, and training each epoch on average takes 0.24min, 0.29min, 0.31min, 0.67min, respectively in our experiments.

The metrics used for measuring similarity between the output images and target ground-truths are structural similarity index measure (SSIM) [23][9] and normalized cross-correlation coefficient [24]. Also, segmentation accuracy between outputs and their corresponding ground-truths is measured by Dice score and False Positive Rate (FPR) [20]. For comparative assessment of the results obtained with the different methods, regarding the different metrics, the average, standard deviation, box plots demonstrating median, percentiles and outliers along with the paired t-test for statistical significance are considered.

Also, the training's loss curves of all folds are drawn and overlaid in a sub-figure to investigate the training progress. Table 1 and Table 2 report the average and standard deviation of the methods for each scenario. The *p-values* of the paired t-test are reported in Table 3 and the pairs which are not significantly different are made bold (p>0.05). More details are given in the subsections that follow.

Table 1: The results achieved by different methods in different Scenarios of tasks.

| | | | **Unet-ST** | **cGAN-ST** | **Unet-MT** | **cGAN-MT** | **Unet-ST** | **cGAN-ST** | **Unet-MT** | **cGAN-MT** |
|---|---|---|---|---|---|---|---|---|---|---|
| **Scen. 1** | **A** | | *Task 1: Convert T2-Flair to T1* | | | | *Task 2: Convert T2-Flair to T1-Inverse* | | | |
| | | **SSIM** | 0.37 ± 0.09 | 0.873 ± 0.05 | 0.272 ± 0.076 | 0.871 ± 0.056 | 0.359 ± 0.079 | 0.941 ± 0.025 | 0.367 ± 0.085 | 0.935 ± 0.030 |
| | | **NCC** | 0.957 ± 0.025 | 0.950 ± 0.028 | 0.958 ± 0.024 | 0.949 ± 0.028 | 0.997 ± 0.001 | 0.997 ± 0.002 | 0.997 ± 0.001 | 0.997 ± 0.002 |
| | | **Epochs** | 280 | 110* | 180 | 115* | 50 | 108* | 180 | 115* |
| | **B** | | *Task 1: Convert T1 to T2-Flair* | | | | *Task 2: Convert T1 to T1-Inverse* | | | |
| | | **SSIM** | 0.452 ± 0.105 | 0.877 ± 0.062 | 0.275 ± 0.065 | 0.872 ± 0.062 | 0.353 ± 0.082 | 0.934 ± 0.030 | 0.366 ± 0.083 | 0.928 ± 0.032 |
| | | **NCC** | 0.937 ± 0.030 | 0.930 ± 0.035 | 0.938 ± 0.028 | 0.928 ± 0.035 | 0.997 ± 0.001 | 0.997 ± 0.001 | 0.998 ± 0.001 | 0.997 ± 0.001 |
| | | **Epochs** | 175 | 90* | 150 | 95* | 49 | 94* | 150 | 95* |
| **Scen. 2** | **A** | | *Task 1: Bias correction on T2-Flair* | | | | *Task 2: Convert biased T2-Flair to T1* | | | |
| | | **SSIM** | 0.325 ± 0.083 | 0.984 ± 0.008 | 0.483 ± 0.097 | 0.993 ± 0.003 | 0.386 ± 0.098 | 0.871 ± 0.056 | 0.372 ± 0.096 | 0.869 ± 0.056 |
| | | **NCC** | 0.963 ± 0.013 | 0.997 ± 0.001 | 0.996 ± 0.003 | 0.997 ± 0.002 | 0.956 ± 0.025 | 0.949 ± 0.028 | 0.957 ± 0.024 | 0.948 ± 0.028 |
| | | **Epochs** | 15 | 13 | 135 | 130 | 280 | 93* | 135 | 130 |
| | **B** | | *Task 1: Bias Correction on T1* | | | | *Task 2: Convert biased T1 to T2-Flair* | | | |
| | | **SSIM** | 0.393 ± 0.091 | 0.983 ± 0.007 | 0.499 ± 0.103 | 0.988 ± 0.008 | 0.442 ± 0.125 | 0.865 ± 0.067 | 0.394 ± 0.089 | 0.869 ± 0.064 |
| | | **NCC** | 0.996 ± 0.002 | 0.996 ± 0.002 | 0.997 ± 0.002 | 0.997 ± 0.003 | 0.935 ± 0.029 | 0.913 ± 0.049 | 0.936 ± 0.032 | 0.923 ± 0.039 |
| | | **Epochs** | 16 | 14 | 104 | 142 | 211 | 79* | 104 | 142 |

*: Discriminator wins for the 10 continuous epochs (force stop)
-: Reached Maximum allowed epochs (force stop)
No sign: reached desired error loss (early stop)

Table 3: Statistical analysis between the results of the methods. See the detail of tasks in Table 1 and 2.

| | | **Task 1** | | | | **Task 2** | | | |
|---|---|---|---|---|---|---|---|---|---|
| | | Unet-ST vs. cGAN-ST | Unet-MT vs. cGAN-MT | Unet-ST vs. Unet-MT | cGAN-ST vs. cGAN-MT | Unet-ST vs. cGAN-ST | Unet-MT vs. cGAN-MT | Unet-ST vs. Unet-MT | cGAN-ST vs. cGAN-MT |
| **Scen. 1** | **A** | 1.162e-47 | 5.2328e-44 | 1.6277e-06 | 0.01019 | **0.38256** | 0.00016243 | 0.041971 | 0.0005788 |
| | **B** | 6.3122e-24 | 7.7839e-30 | **0.1011** | 2.8427e-05 | 3.4878e-09 | 3.5696e-26 | 8.9873e-45 | 1.0541e-19 |
| **Scen. 2** | **A** | 1.657e-156 | 2.2753e-24 | 9.46e-155 | 0.022767 | 8.0895e-59 | 3.073e-52 | 0.0002652 | 5.2617e-05 |
| | **B** | 0.00151 | 1.9444e-06 | 5.8674e-30 | **0.27057** | 1.9103e-25 | 5.0658e-39 | **0.12454** | 2.0078e-11 |
| **Scen. 3** | **A** | 1.672e-204 | 6.0199e-111 | 6.9416e-160 | **0.15509** | 3.7322e-106 | 2.2225e-305 | 1.1697e-12 | 2.2644e-298 |
| | **B** | 3.6247e-57 | 8.1886e-56 | 1.2244e-27 | **0.071969** | 2.2465e-20 | 2.7326e-156 | 8.346e-29 | 1.1111e-222 |
| **Scen. 4** | **A** | 4.6354e-185 | 4.803e-252 | 1.244e-147 | 1.1677e-13 | 4.284e-224 | 1.2408e-132 | 3.1934e-226 | 2.3944e-124 |
| | **B** | 4.731e-64 | 1.653e-188 | 6.149e-09 | 2.6827e-87 | 8.8715e-137 | 3.7491e-226 | 7.5824e-143 | 4.236e-219 |
| **Scen. 5** | **A** | 4.6354e-185 | 7.1581e-127 | 1.6173e-131 | 2.7817e-25 | 1.5597e-145 | 5.4444e-170 | 0.011016 | 2.0594e-154 |
| | **B** | 4.731e-64 | 1.4054e-131 | 1.767e-05 | 9.7532e-16 | 7.2349e-44 | 1.0895e-189 | 8.401e-22 | 3.3514e-186 |

Segmentation tasks are based on Dice scores, and the rest are based on NCC.

Table 2: The results achieved by different methods in different Scenarios of tasks.

| | | | Unet-ST | cGAN-ST | Unet-MT | cGAN-MT | | Unet-ST | cGAN-ST | Unet-MT | cGAN-MT |
|---|---|---|---|---|---|---|---|---|---|---|---|
| **Scen. 3** | **A** | *Task 1: Segmentation on T2-Flair* | | | | | *Task 2: Convert T2-Flair to T1* | | | | |
| | | **Dice** | 0.520 ± 0.091 | 0.789 ± 0.027 | 0.727 ± 0.045 | 0.788 ± 0.023 | **SSIM** | 0.442 ± 0.054 | 0.911 ± 0.027 | 0.500 ± 0.082 | 0.744 ± 0.045 |
| | | **FPR** | 0.040 ± 0.017 | 0.174 ± 0.037 | 0.075 ± 0.020 | 0.167 ± 0.030 | **NCC** | 0.982 ± 0.005 | 0.979 ± 0.006 | 0.983 ± 0.005 | 0.950 ± 0.008 |
| | | **Epoch** | 500- | 42* | 500- | 55* | | 425 | 89* | 500- | 55* |
| | **B** | *Task 1: Segmentation on T1* | | | | | *Task 2: Convert T1 to T2-Flair* | | | | |
| | | **Dice** | 0.774 ± 0.044 | 0.818 ± 0.021 | 0.699 ± 0.129 | 0.819 ± 0.020 | **SSIM** | 0.431 ± 0.076 | 0.930 ± 0.021 | 0.480 ± 0.181 | 0.745 ± 0.042 |
| | | **FPR** | 0.068 ± 0.025 | 0.149 ± 0.028 | 0.512 ± 0.946 | 0.156 ± 0.030 | **NCC** | 0.977 ± 0.010 | 0.974 ± 0.012 | 0.970 ± 0.015 | 0.931 ± 0.018 |
| | | **Epoch** | 500- | 46* | 500- | 49* | | 251 | 87 | 500- | 49* |
| **Scen. 4** | **A** | *Task 1: Segmentation on biased T2-Flair* | | | | | *Task 2: Bias correction on T2-Flair* | | | | |
| | | **Dice** | 0.495 ± 0.108 | 0.797 ± 0.020 | 0.129 ± 0.159 | 0.783 ± 0.028 | **SSIM** | 0.302 ± 0.043 | 0.987 ± 0.004 | 0.674 ± 0.154 | 0.832 ± 0.036 |
| | | **FPR** | 0.061 ± 0.059 | 0.174 ± 0.031 | 0.006 ± 0.007 | 0.185 ± 0.041 | **NCC** | 0.947 ± 0.015 | 0.998 ± 0.001 | 0.999 ± 0.001 | 0.972 ± 0.014 |
| | | **Epoch** | 500- | 48* | 500- | 53* | | 39 | 14 | 500- | 53* |
| | **B** | *Task 1: Segmentation on biased T1* | | | | | *Task 2: Bias correction on T1* | | | | |
| | | **Dice** | 0.672 ± 0.140 | 0.815 ± 0.018 | 0.628 ± 0.061 | 0.800 ± 0.022 | **SSIM** | 0.376 ± 0.053' | 0.989 ± 0.003 | 0.625 ± 0.098 | 0.793 ± 0.045 |
| | | **FPR** | 0.083 ± 0.091 | 0.165 ± 0.033 | 0.046 ± 0.018 | 0.163 ± 0.034 | **NCC** | 0.987 ± 0.006 | 0.998 ± 0.000 | 0.999 ± 0.001 | 0.975 ± 0.007 |
| | | **Epoch** | 500- | 50* | 500- | 52* | | 63 | 11 | 500- | 52* |
| **Scen. 5** | **A** | *Task 1: Segmentation on biased T2-Flair* | | | | | *Task 2: Convert biased T2-Flair to T1* | | | | |
| | | **Dice** | 0.495 ± 0.108 | 0.797 ± 0.020 | 0.723 ± 0.038 | 0.788 ± 0.026 | **SSIM** | 0.444 ± 0.056 | 0.906 ± 0.025 | 0.512 ± 0.077 | 0.752 ± 0.052 |
| | | **FPR** | 0.061 ± 0.059 | 0.174 ± 0.031 | 0.083 ± 0.019 | 0.182 ± 0.042 | **NCC** | 0.981 ± 0.006 | 0.976 ± 0.006 | 0.981 ± 0.006 | 0.948 ± 0.014 |
| | | **Epoch** | 500- | 48* | 500- | 64* | | 500- | 38 | 500- | 64* |
| | **B** | *Task 1: Segmentation on biased T1* | | | | | *Task 2: Convert biased T1 to T2-Flair* | | | | |
| | | **Dice** | 0.672 ± 0.140 | 0.815 ± 0.018 | 0.709 ± 0.064 | 0.805 ± 0.022 | **SSIM** | 0.437 ± 0.066 | 0.922 ± 0.023 | 0.509 ± 0.143 | 0.741 ± 0.046 |
| | | **FPR** | 0.083 ± 0.091 | 0.165 ± 0.03 | 0.046 ± 0.021 | 0.161 ± 0.037 | **NCC** | 0.974 ± 0.012 | 0.969 ± 0.015 | 0.973 ± 0.012 | 0.922 ± 0.023 |
| | | **Epoch** | 500- | 50* | 500- | 49* | | 68 | 12 | 500- | 49* |

*: Discriminator wins for the 10 continuous epochs (force stop)
-: Reached Maximum allowed epochs (force stop)
No sign: reached desired error loss (early stop)

### 1) *Scenario 1: Cross-modality conversion*

The objective of scenario 1 is to find out whether or not a multitasking approach can be beneficial for cross-modality tasks. This scenario includes two sub-scenarios: **A)** Input: T2-Flair, Task 1: Conversion to T1, Task 2: Conversion to T1-Inverse. **B)** Input: T1, Task 1: Conversion to T2-Flair, Task 2: Conversion to T1-Inverse. The achieved results for these experiments are shown in Figure 1, in which the left and right sides are for experiments A and B respectively. As can be seen, there are no significant differences between the single-task and the multitask approach for both U-net and cGAN. Regarding the NCC score, the accuracy of the methods is almost the same, but there is a huge difference between U-net and cGAN according to the SSIM score because Unet was not able to preserve the intensities precisely, while it preserved the shapes well. Considering the loss curves, in the single-task approach, task 1 seems to have a harder time reaching the threshold loss of 0.01. The case is true for both scenarios of A and B; however, it seems that scenario A is a harder problem to solve and it requires a greater number of epochs for both tasks to be trained. In both of these scenarios, the multitasking approach helped Task 1 by reducing the number of epochs to converge, speeding up the training time as a consequence. Additionally, since multitasking does not affect accuracy, performing both tasks while minimizing network parameters is possible.

#### 2) *SCENARIO 2: Bias correction & cross-modality conversion*

This scenario 2 investigates whether cross-modality task and bias correction task can help each other. This scenario includes two sub-scenarios: A) Input: Contaminated T2-Flair, Task 1: Bias correction, Task 2: Conversion to T1; B) Input: Contaminated T1, Task 1: Bias correction, Task 2: Conversion to T2-Flair. The results of these experiments are shown in Figure 2, where the left and right sides correspond to experiments A and B, respectively. In these tasks, almost all the methods work well, but Unet is slightly better than the other. Loss curves show that bias correction is the less challenging task which tends to converge in fewer epochs. Since multitasking did not lead to better accuracy, it would be reasonable if both tasks were of interest to us. On the other hand, if cross-modality conversion is our desired task, using multititasking would be beneficial according to the faster and better convergence for Unet and cGAN, respectively. It should be emphasized that only cGAN in the second task faced forced stop, all other experiments reached the desired loss value.

#### 3) *SCENARIO 3: Tissue segmentation & cross-modality conversion*

The aim in this scenario is to investigate whether the segmentation task and cross-modality conversion task help each other. This scenario includes two sub-scenarios: **A)** Input: T2-Flair, Task 1: Segmentation, Task 2: Conversion to T1; **B)** Input: T1, Task 1: Segmentation, Task 2: Conversion to T2-Flair. The achieved results for these experiments are shown in Figure 3 in which the left and right sides are for experiments A and B, respectively. As shown in this figure, the cGAN was significantly better than Unet in the segmentation task. Multitasking is beneficial to segmentation tasks (especially for Unet) as confirmed in other studies, but it is unfavorable for the conversion task. In the training phase, both cGAN and Unet can yield a desired loss value for the second task, but faced force stop for segmentation task for either single task or multitask, which underlines the difficulties of the segmentation task.

#### 4) *SCENARIO 4: Segmentation on contaminated images and Bias correction*

This scenario seeks to explore whether segmentation and bias correction can help each other. This scenario includes two sub-scenarios: A) Input: Contaminated T2-Flair, Task1: Segmentation, Task 2: Bias correction; B) Input: Contaminated T1, Task 1: Segmentation, Task 2: Bias correction. Results are shown in Figure 4. As expected, segmentation accuracy decreases in comparison to scenario 3. The cGAN method is significantly better than Unet in both single tasks. In addition, multitasking is not helpful for segmentation task and for the bias correction of cGAN. The Unet's bias correction task could have benefitted from multitasking, but this would have resulted in greater computation cost and time. As before, experiments involving segmentation encountered a force stop.

#### *5) SCENARIO 5: Segmentation on contaminated image and cross-modality conversion*

This scenario aims to investigate whether segmentation and cross-modality conversion can help each other when dealing with contaminated images. This scenario includes two sub-scenarios: **A)** Input: Contaminated T2-Flair, Task 1: Segmentation, Task 2: Conversion to T1; **B)** Input: Contaminated T1, Task 1: Segmentation, Task 2: Conversion to T2-Flair. The results are shown in Figure 5. In this scenario, in contrast to scenario 4, multitasking is helpful for Unet in the segmentation task, but it is not so advantageous for cGAN; and as before, the experiments with segmentation task cannot reach the desired loss value in the training phase and faced force stop as well.

#### *6) Impact of the Hyperparameters*

In this subsection, the effect of some hyper-parameters for the cGAN method are shown including slice size (256x256 vs. 512x512), kernel size of the filter (3x3 vs. 4x4), and L1-weight (1 vs. 10 vs. 30). Figure 6 illustrates and compares these different settings for the cross-modality conversion task and shows that setting A is the best which was used in our experiments. The impact of image size and kernel size is in the details and edges of the tissues and the impact of the l1-weight is on the rate of convergence.

## *D. CONCLUSION*

In this report, five scenarios of multitasking in neuroimaging analysis were considered. These different scenarios evaluated the merits of multitasking on tissue segmentation, cross-modality conversion, and bias correction. In general, bias correction and cross-modality conversion tasks are easier for neural networks to handle, and the learning methods are able to scale to the desired loss within a short period of time. Multitasking the two tasks does not result in significant improvements in accuracy but does provide the means to reduce the network parameters. On the other hand, the tissue segmentation task is a more complicated problem, and most of the time the training fails to achieve the desired loss value. It means, if our aim task is bias correction or cross-modal conversion then multitasking with the segmentation task is not reasonable, as it makes the system converge even later with even less accuracy. However, if tissue segmentation is the main objective, then multitasking with cross-modality conversion is beneficial.

## *E. REFERENCES*

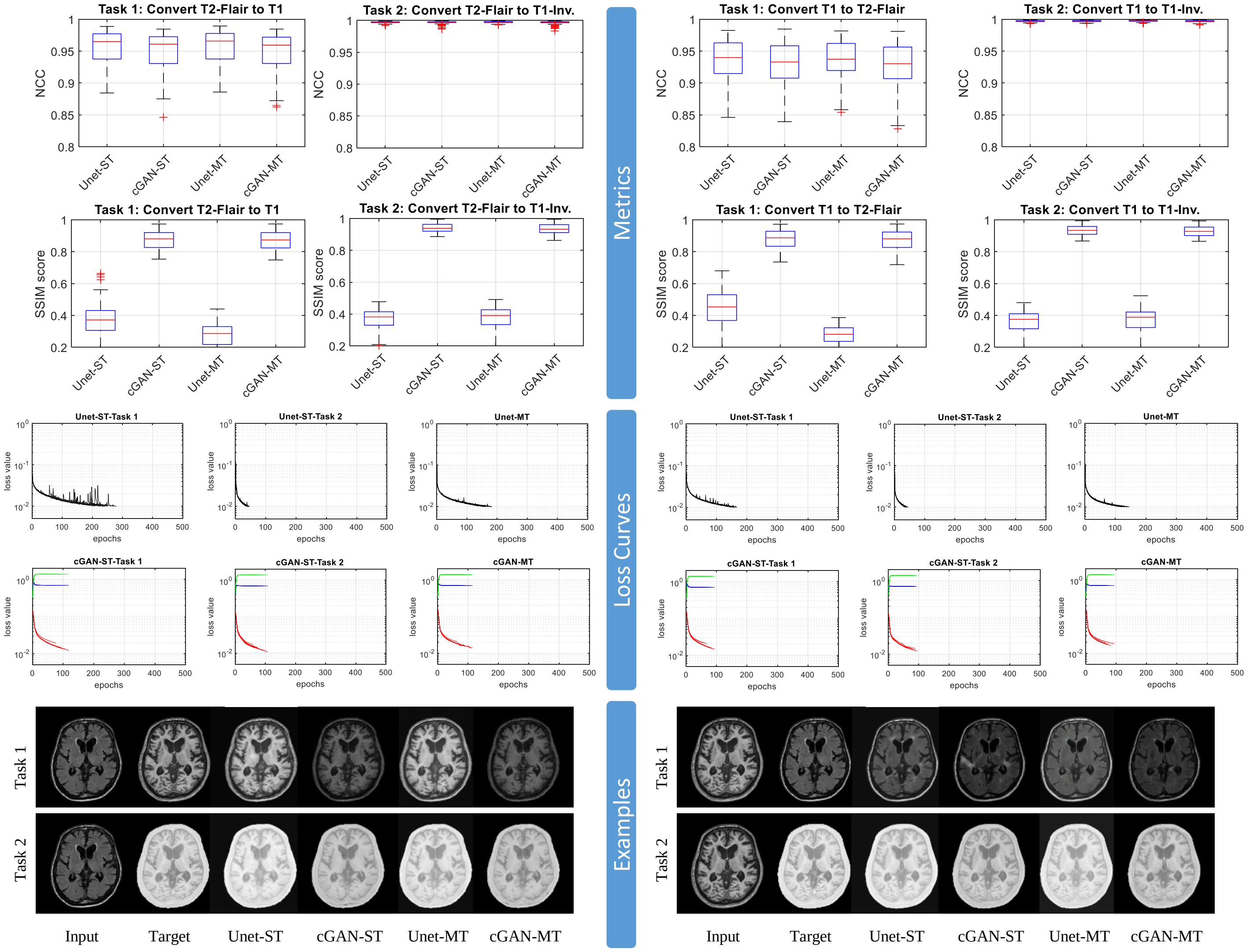

Figure 1: Results for scenario 1: cross-modality conversion. Left and right are sub-scenarios A and B. Up to the bottom are boxplots, loss curves, and an exemplary slice.

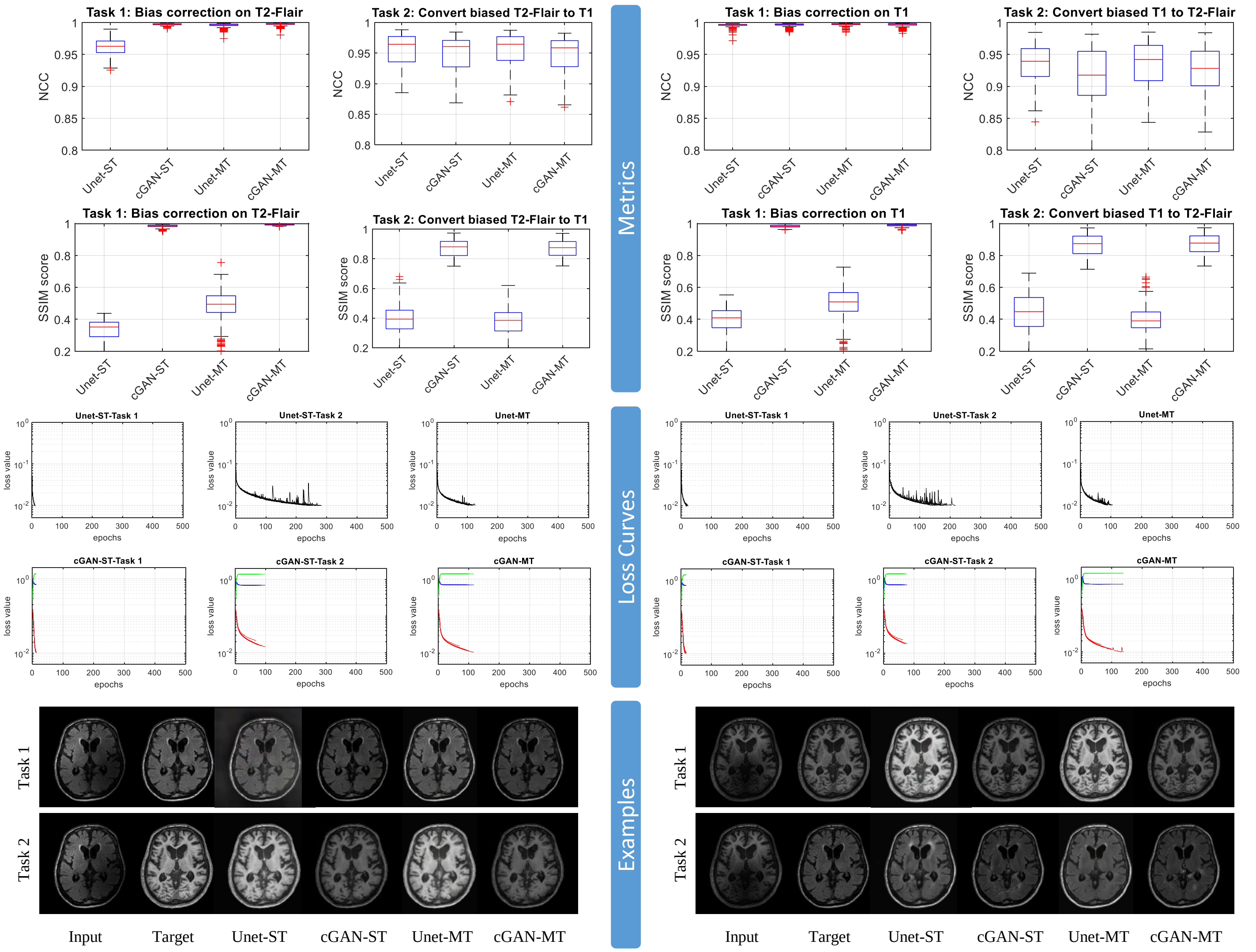

Figure 2: Results for Scenario 2: bias correction and cross-modality conversion for the contaminated image. Left and right are the sub-scenarios A and B. Up to bottom are boxplots, loss curves and exemplary slice.

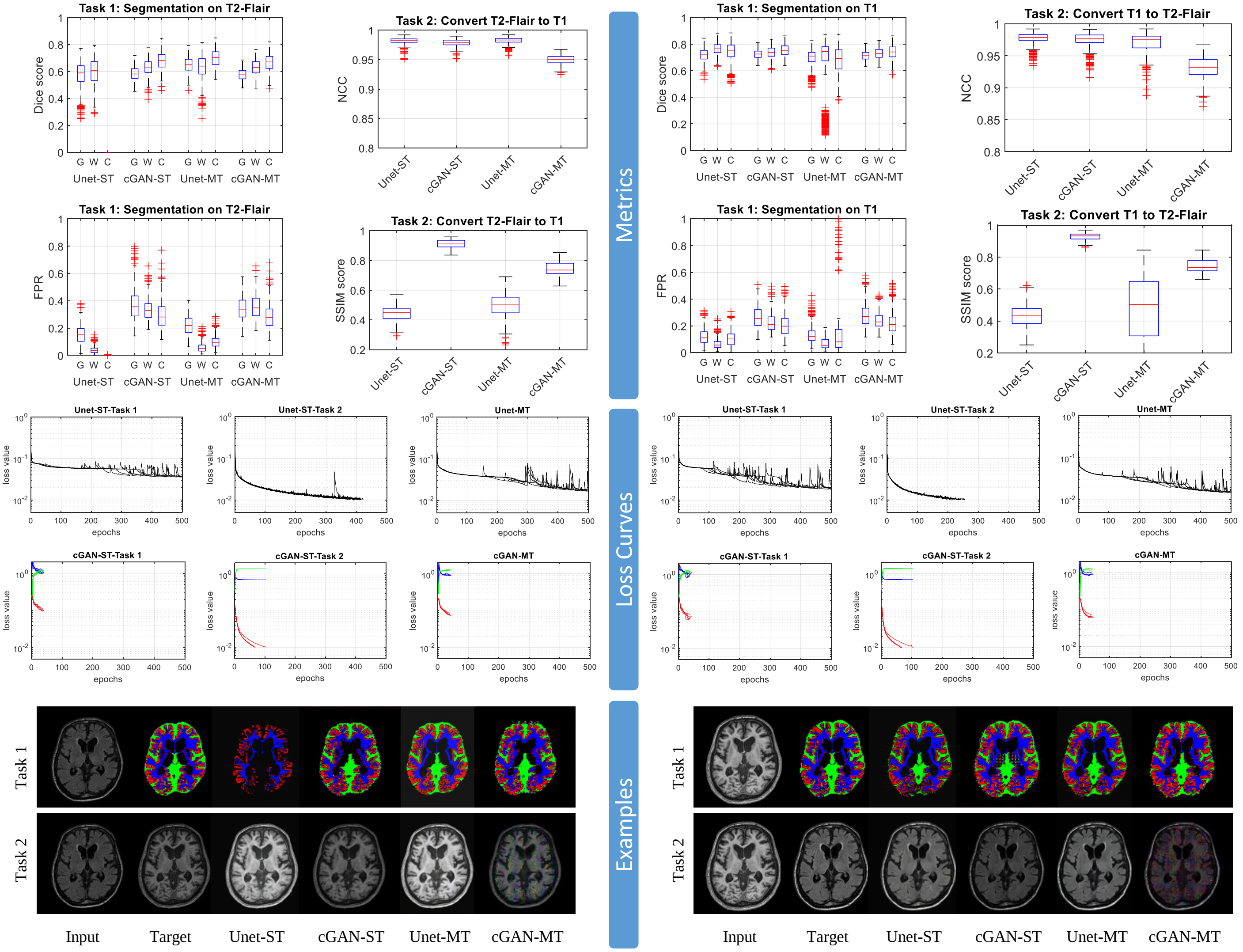

Figure 3: Results for Scenario 3: tissue segmentation and cross-modality conversion. Left and right are sub-scenarios A and B. Up to bottom are boxplots, loss curves and exemplary slice. G, W, C denote gray matter, white matter, CSF shown in red, blue and green masks, respectively.

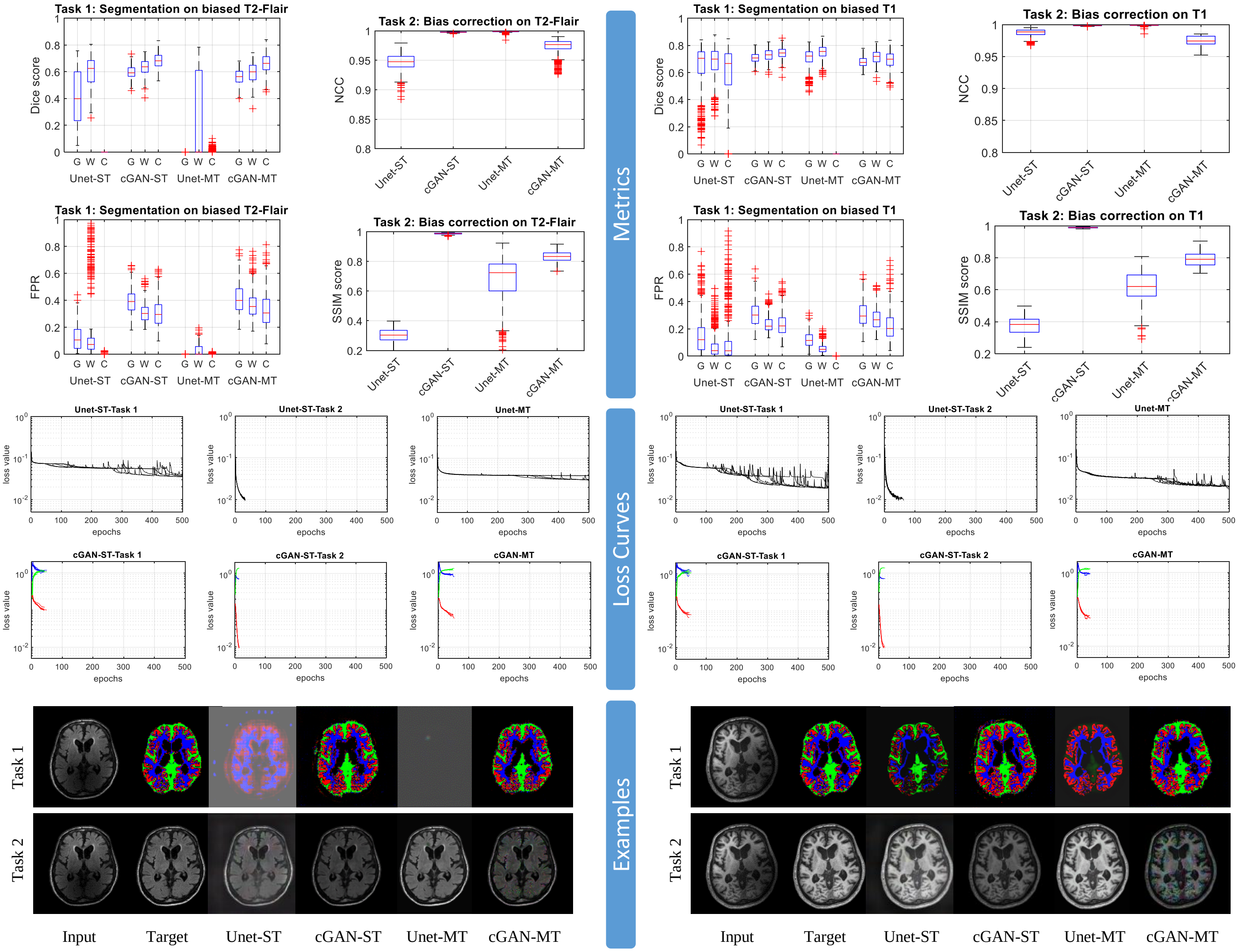

Figure 4: Results for scenario 4: tissue segmentation and bias segmentation. Left and right are sub-scenarios A and B. Up to bottom are boxplots, loss curves and exemplary slice. G, W, C stands for gray matter, white matter, CSF and are shown with red, blue and green masks, respectively.

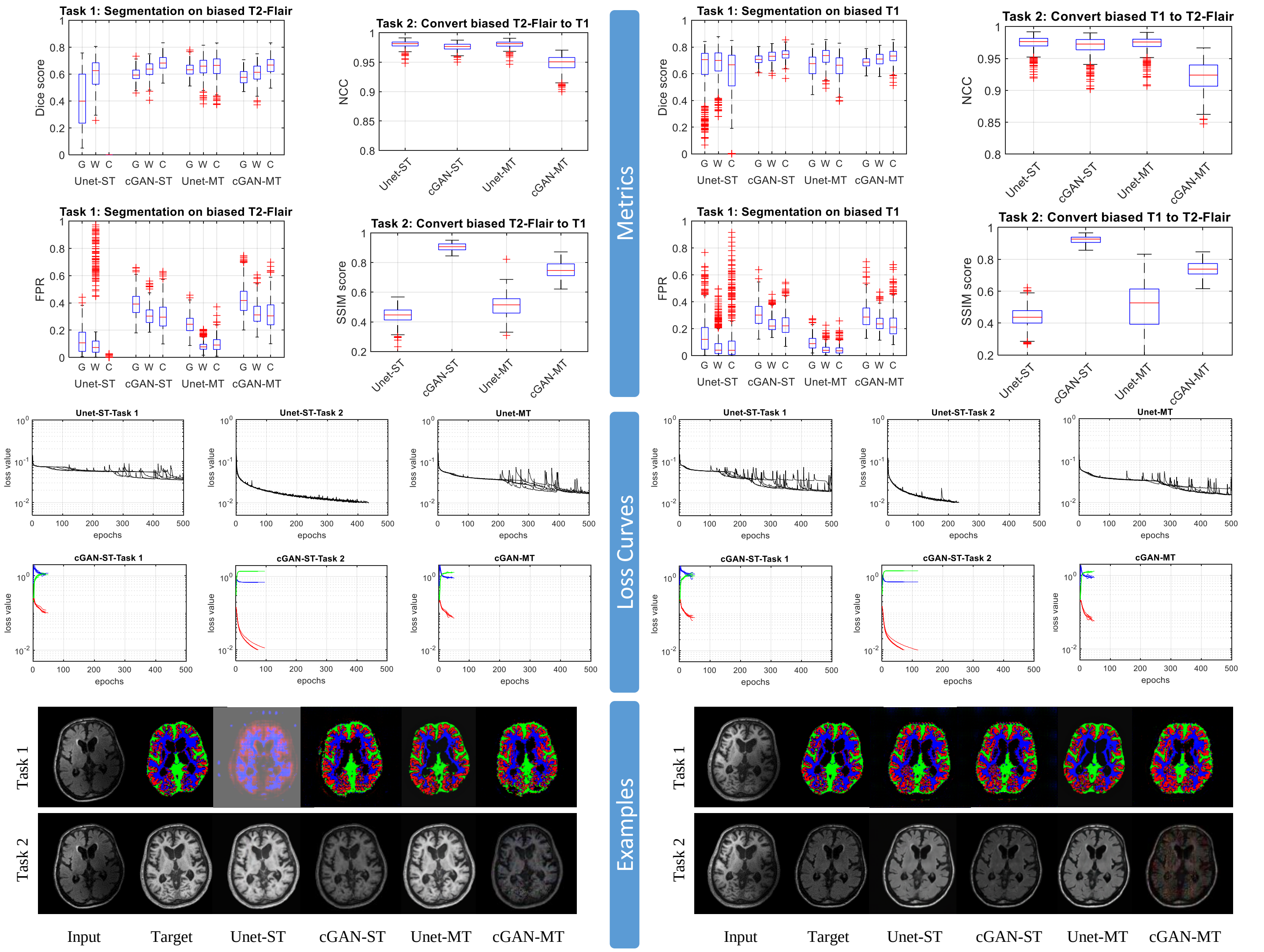

Figure 5: Results for scenario 5: tissue segmentation and bias segmentation. Left and right are sub-scenarios A and B. Up to bottom are boxplots, loss curves and exemplary slice. G, W, C stands for gray matter, white matter, CSF and are shown with red, blue and green masks, respectively.

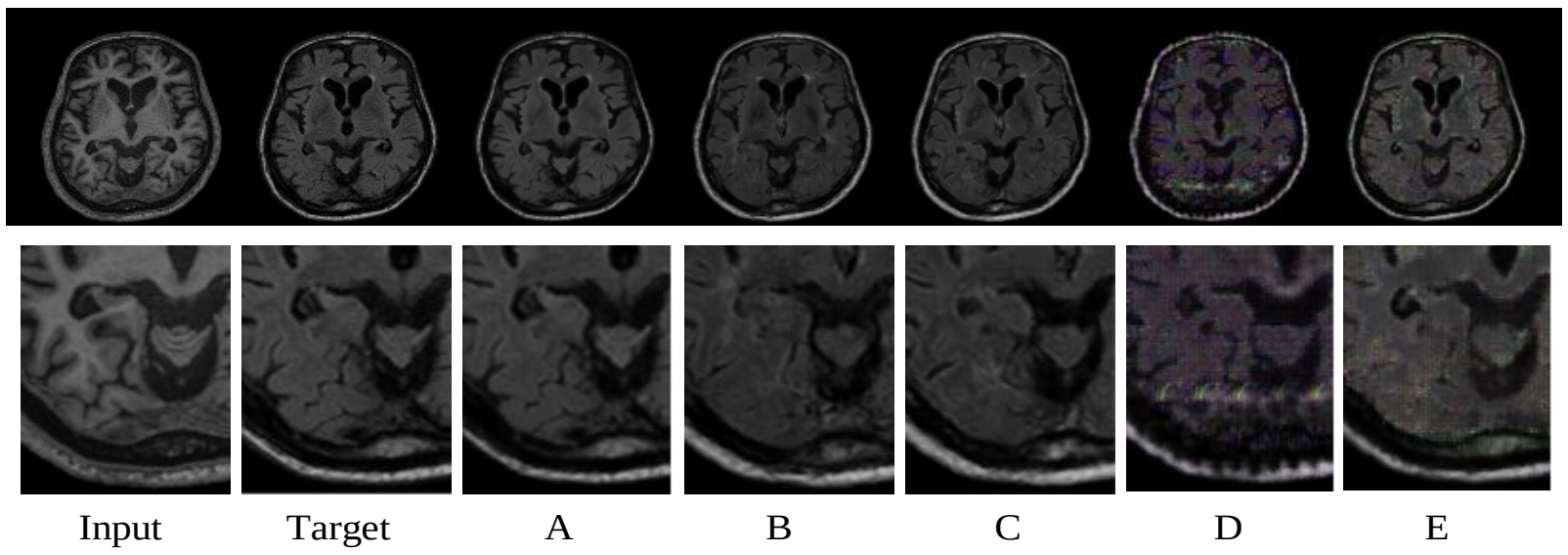

Figure 6: Illustrating the impact of hyper-parameters for the cGAN-ST method in different settings. A) l1-wirght: 10, kernel size: 3x3, slice size: 512x512, Adam optimizer with learning rate 0.0002, batch size of 20; B) Setting A but with slice size 256x256; C) Setting A but with kernel size 4x4; D) Setting A with l1-weight equal to 1, E) Setting A with l1-weight equal to 30.